\documentclass[aps,prl,twocolumn,groupedaddress,showpacs]{revtex4-1}
\usepackage{graphicx}
\usepackage{psfrag}
\def\be{\begin{equation}}
\def\ee{\end{equation}}

\begin{document}
\date{\today}

\title{Ergodicity Breaking and Parametric Resonances in Systems with Long-Range Interactions}
\author{Fernanda P. da C. Benetti}
\author{Tarc\'isio N. Teles}
\author{Renato Pakter}
\author{Yan Levin}
\affiliation{Instituto de F\'{\i}sica, Universidade Federal do Rio Grande do Sul, Caixa Postal 15051, CEP 91501-970,
Porto Alegre, RS, Brazil}

\begin{abstract}

We explore the mechanism responsible for the  ergodicity breaking in systems with long-range
forces.  In thermodynamic limit such systems do not evolve to the Boltzmann-Gibbs 
equilibrium, but become trapped in an out-of-equilibrium quasi-stationary-state.  Nevertheless, we show
that if the initial distribution satisfies a specific 
constraint --- a generalized virial condition --- the quasi-stationary-state is very 
close to ergodic and can be
described by Lynden-Bell statistics.  On the other hand if the generalized virial condition is violated, 
parametric resonances are excited, leading to chaos and ergodicity breaking.  

\end{abstract}


\maketitle

Statistical mechanics of systems in which particles interact through long-ranged potentials is
fundamentally different from the statistical mechanics of systems with short-range forces~\cite{Campa09}.  
In the latter case,
starting from an arbitrary initial condition (microcanonical
ensemble) systems evolve to a thermodynamic equilibrium in which particle distribution functions are given
by the usual Boltzmann-Gibbs statistical mechanics~\cite{Gibbs}. The state of thermodynamic 
equilibrium does not depend on the specifics of the initial distribution, but only on the
global conserved quantities such as energy, momentum, angular momentum, etc.  The situation is very different
for systems in which particles interact through long-range potentials, such as gravity or unscreened 
Coulomb interactions~\cite{Levinprl,TaLe10,JoWo,Tel2011}.
In this case, it has been observed in numerous simulations  that these systems do not relax to thermodynamic equilibrium,
but become trapped in a quasi-stationary state (qSS), the lifetime of which diverges with the number of 
particles~\cite{Barre2001,Kav2007,Tel2011,TaLe10}.
The distribution functions in this quasi-stationary state do not obey the Boltzmann-Gibbs statistical mechanics ---
and in particular,  particle velocities do not follow the Maxwell-Boltzmann distribution, but  
depend explicitly on the initial condition.
It has been an outstanding challenge of statistical mechanics to quantitatively predict the final stationary state 
reached
by systems with unscreened long-range forces,  without having to explicitly solve the $N$-body dynamics or  
the collisionless Boltzmann (Vlasov) equation.

Some $40$ years ago Lynden-Bell (LB) proposed a generalization of the Boltzmann-Gibbs statistical mechanics 
to treat systems with long-range interactions~\cite{Ly67}. 
Lynden-Bell's construction was based on the Boltzmann counting, but instead of using particles, LB
worked directly with the levels of the distribution function.  The motivation for 
this approach was the observation that dynamical
evolution of the distribution function for systems with long-range interactions is governed by the Vlasov equation~\cite{Br77}.
This equation has an infinite number of conserved quantities, Casimirs --- any local functional
of the distribution function is a Casimir invariant of the Vlasov dynamics. In particular if the initial
distribution function is discretized into levels, the volume of each level must be preserved by the Vlasov flow.
For an initially one-level distribution function, Vlasov dynamics requires that 
the phase space density does not exceed that of the initial distribution --- one-particle 
distribution function over the reduced phase space ($\mu$-space) evolves as an incompressible fluid.   
Using this constraint in a combination
with the Boltzmann counting, LB was able to derive a coarse-grained entropy, the maximum of which he
argued should correspond to the most-probable distribution  --- the one that should describe the equilibrium state.
Numerous simulations, however, showed that, in general, 
Lynden-Bell statistics was not able to account for the particle distribution in self-gravitating systems, and
the theory has been abandoned in the astrophysical context.  Recently, however, Lynden-Bell's work has been rediscovered
by the Statistical Mechanics community, which showed that for some systems, specifically the widely studied
Hamiltonian Mean Field Model (HMF), Lynden-Bell's approach could make reasonable predictions about the structure
of the phase diagram~\cite{AnCa07}.  The fundamental question that needs to be addressed is: Under what conditions
can Lynden-Bell statistics be used to accurately describe systems with long-range interactions?
This will be the topic of the present Letter.

To be specific, we will study the HMF model~\cite{Campa09}, which has become a test
bench for theories of systems with long-range forces. However, our results and methods are  
completely general and can be applied to other systems, 
such as self-gravitating clusters
or confined non-neutral plasmas.  
The HMF model consists of $N$ particles restricted to move on a circle of radius one.  The dynamics 
is governed by the Hamiltonian
\be
H=\sum_{i=1}^N {p_i^2\over 2}+{1\over 2 N}\sum _{i,j=1}^N [1-\cos(\theta_i-\theta_j)],
\ee
where the angle $\theta_i$ is the position of  $i$'th particle
and $p_i$ is its conjugate momentum~\cite{AnCa07,AnFa07,AnFa07b}.
The {\it macroscopic} behavior of the system is characterized by the magnetization vector
${\bf M}=(M_x,M_y)$, where $M_x\equiv \langle \cos\theta \rangle$, 
$M_y\equiv \langle \sin\theta \rangle$, and $\langle\cdots\rangle$ stands for the average
over all  particles.  The Hamilton's equations of motion for each particle reduce to  
\be
\ddot\theta_i=-M_x(t)\sin\theta_i(t)+M_y(t)\cos \theta_i(t).
\label{evol}
\ee
Since the Hamiltonian does not have explicit time dependence, the average energy per particle,
\be
u={H\over N}={\langle p^2\rangle\over 2}+{1-M(t)^2\over 2}\,,
\label{energyu}
\ee 
is conserved. 

The failure of  LB  theory in the astrophysical context was attributed to  
incomplete relaxation, lack of good mixing, or broken
ergodicity~\cite{Muka2005}.  The mechanisms behind this failure have not been elucidated.  On the other hand, it has been recently
observed
that if the initial distribution is virialized --- satisfies the virial condition --- LB's approach was able to quite accurately 
predict the stationary state of gravitational and Coulomb systems \cite{Levinprl,TaLe10,JoWo,Tel2011}.  Unfortunately, the virial theorem can be derived
only for potentials which are homogeneous functions.  This is not the case for the HMF model.  Nevertheless, the fact that LB theory
seems to apply under some  conditions makes one wonder if such conditions can be found for arbitrary long-range potentials,
which are not in general homogeneous functions.

To answer the questions posed above, we note that if the initial distribution is virialized, macroscopic
oscillations of observables should be diminished.  On the other hand, if the system is far from virial, the mean-field
potential that each particle feels will undergo strong oscillations.  It is then possible for some
particles to enter in resonance with the oscillations of the mean-field, gaining large amounts of energy.  The parametric
resonances will result in the occupation of regions of the phase-space which are highly improbable, from the point
of view of Boltzmann-Gibbs or LB statistics~\cite{BaCh08}.  Furthermore, resonant particles will take away energy from
collective oscillations producing  a form of non-linear Landau damping~\cite{La46}. After some time, macroscopic oscillations
will die out and  each particle will feel  only the static mean-field potential.  From that point on, particle dynamics
will become completely regular, with no energy exchange possible between the different particles.  The particles
which have gained a lot of energy from the parametric resonances will  be trapped forever in the  highly improbable 
regions of the phase-space,  unable  
to thermalize with the rest of the system.  

To see how the theoretical picture advocated above can be applied to the HMF, we first
derive a generalized virial condition for this model.   
For simplicity  
we will consider initial distributions of the ``water-bag'' form in $(\theta,p)$. 
Without loss of
generality, we choose a frame of reference where $\langle \theta \rangle =0$ and 
$\langle p \rangle =0$. The one-particle initial distribution function then reads
\be
f_0(\theta,p)={1\over 4\theta_0 p_0} \Theta (\theta_0-|\theta|)\, \Theta (p_0-|p|),
\label{f0}
\ee
where $\Theta$ is the Heaviside step function, and $|\theta_0|$ and $|p_0|$ are the maximum  values of angle and 
momentum, respectively. Note that from symmetry, $M_y(t)=0$ at all times.
When the dynamics starts, the mean-squared particle position will evolve with time.  We define
the envelope of the particle distribution as  $\theta_e(t) = \sqrt{3\langle\theta^2\rangle}$, so that
$\theta_e(0)=\theta_0$.   We next differentiate $\theta_e (t)$ twice with respect to time to obtain the 
envelope equation of motion,
\begin{equation}\label{eq:envelop}
\ddot{\theta_e}=\frac{3\langle p^2\rangle}{\theta_e}+\frac{3\langle\theta\ddot{\theta}\rangle}{\theta_e}-\frac{9\langle\theta p\rangle^2}{\theta_e^3}.
\end{equation}
Using the conservation of energy, $\langle p^2\rangle=2u+M_x^2(t)-1$.
To calculate, $\langle\theta\ddot{\theta}\rangle$, we use the  equation of motion for $\theta$.  Supposing that  
the distribution of angles remains close to uniform on the interval $[-\theta_e,\theta_e]$, we obtain
\begin{eqnarray}
\langle\theta\ddot{\theta}\rangle &=& \frac{-M_x(t)}{2\theta_e} \int_{-\theta_e(t)}^{\theta_e(t)} \theta\sin \theta d\theta \nonumber \\
 &=& M_x(t)\cos\theta_e(t)-M_x^2(t),
\end{eqnarray}
where the magnetization $M_x(t)$ is
\begin{eqnarray}\label{mag}
M_x(t) &=&\frac{1}{2\theta_e} \int^{\theta_e(t)}_{-\theta_e(t)} d\theta \cos\theta \nonumber \\
 &=& \frac{\sin\theta_e(t)}{\theta_e(t)}.
\end{eqnarray}
Neglecting the correlations between positions and velocities, $\langle\theta p\rangle =0$, we finally obtain 
a dynamical equation for the envelope
\begin{eqnarray}\label{env}
\ddot{\theta_e}=\frac{3}{\theta_e(t)}\left(2u+M_x(t)\cos\theta_e(t)-1\right),
\end{eqnarray}
where  $u=p_0^2/6+(1-M_0^2)/2$ and 
$M_0= \sin(\theta_0)/\theta_0$. The generalized virial 
condition is defined by the stationary envelope, $\ddot{\theta_e}=0$, which means that along
the curve
\begin{eqnarray}
\label{curveLB}
(2u-1)\theta_0+\sin\theta_0 \cos\theta_0 =0.
\end{eqnarray}
magnetization remains approximately invariant.
In Fig. 1 we plot Eq.(\ref{curveLB}) in the $M_0-u$ plane and compare it with the 
full molecular dynamics simulation of the HMF model.  As can be
seen, agreement between the theory and simulation is excellent.
\begin{figure} 
\includegraphics[scale=0.8,width=9cm]{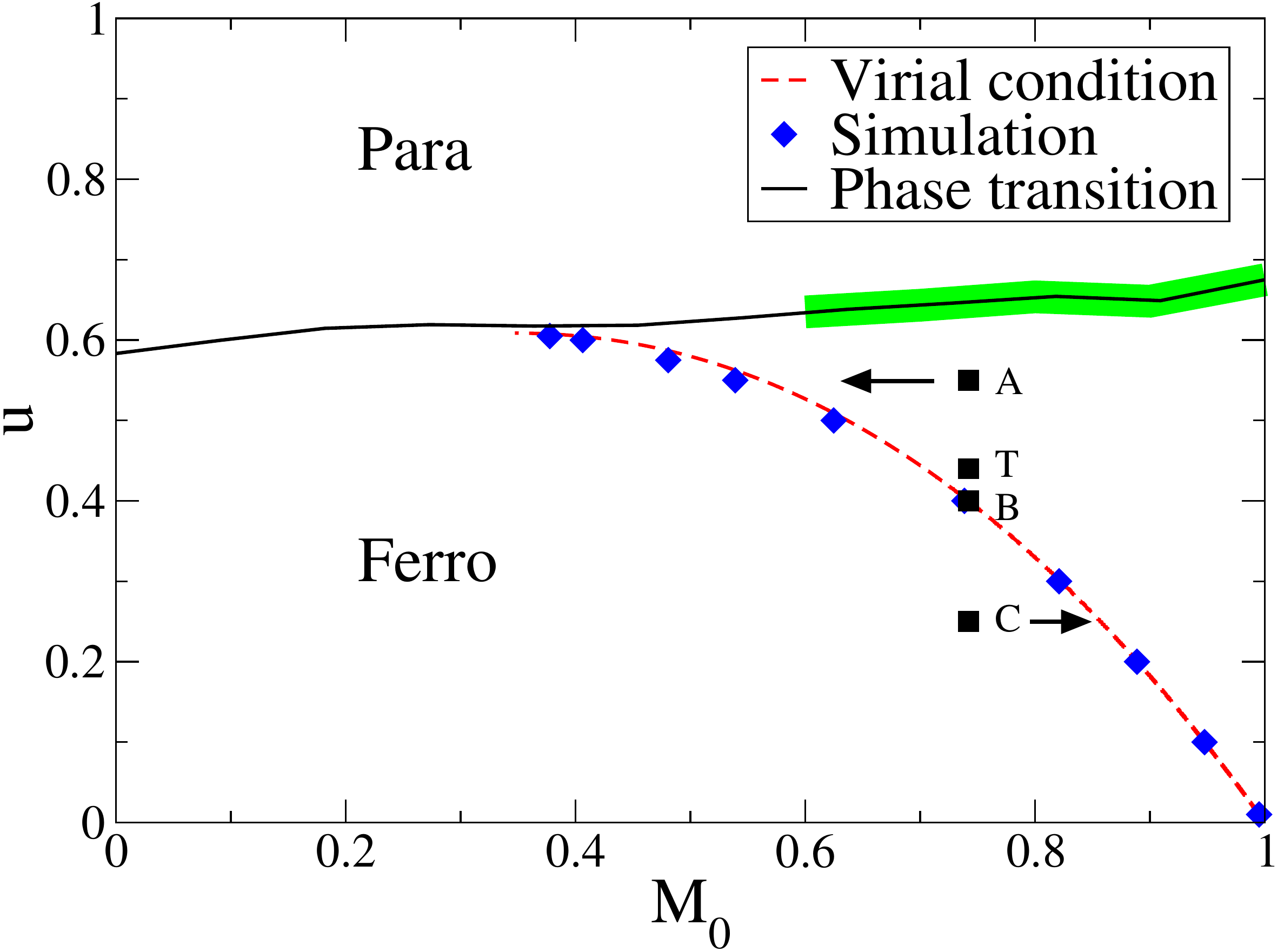}
\caption{Phase diagram of the HMF model obtained using the molecular dynamics simulations.  
Solid curve is the line of first order phase transitions separating paramagnetic and
ferromagnetic phases. This line  extends up to $M_0=0.6$, after
which point the order of the phase transition, shaded region (green line), becomes unclear, with 
strong dependence on the initial conditions and 
various reentrant transitions  occurring in this region.
Dashed curve is the generalized virial condition, Eq.(\ref{curveLB}). Along this curve oscillations of 
the envelope are suppressed.  Diamonds are the results of  simulation. 
Starting with the initial energy and magnetization along the
virial curve, diamonds show the final magnetization to which the system relaxes. 
For points along this curve,  
the final magnetization is almost identical to the initial one. Note that the generalized 
virial curve terminates at $M_0=0.34$ slightly below the phase transition line.  This small difference, however, is sufficient to
invalidate the Lynden-Bell theory, which for $M_0=0.4$ predicts a second order phase transition, while the simulations show
that the phase transition is of first order \cite{PaLe11}.  Points (A), (B), and (C) correspond to the initial conditions for 
the distribution 
functions shown in Fig. 2. The Poincar\'e sections of the test particle dynamics for the initial conditions described by  
the points (B) and (T) are shown in Fig. 3. Finally, we note that since the  stationary distribution must satisfy the 
virial condition and the energy is conserved, 
Eq.(\ref{curveLB}) allows us to predict the magnetization to which the system will evolve for initial conditions lying inside the
ferromagnetic region, see the arrows for points (A) and
(C).}
\label{fig1}
\end{figure}

Along the generalized virial condition curve, Eq. (\ref{curveLB}), 
the magnetization --- and, therefore, the mean-field potential acting on each particle --- of the HMF model has only  
microscopic oscillations and the 
parametric resonances
are suppressed. Under these conditions, we expect that LB theory will be valid.   
The coarse grained entropy within the LB approach is given by 
\begin{equation}
s(f)=-\int dp d\theta \left[\frac{f}{\eta_0}\ln\frac{f}{\eta_0}+\left(1-\frac{f}{\eta_0}\right)\ln\left(1-\frac{f}{\eta_0}\right)\right],
\end{equation}
where $\eta_0=1/4 \theta_0 p_0$~\cite{StCh09}.  Maximizing this entropy under the constraints of energy and particle conservation,
we obtain the equilibrium distribution function 
\begin{equation}
f(p,\theta)=\frac{\eta_0}{e^{\beta\left[\frac{p^2}{2}- M_x\cos\theta-\mu\right]+1}}.
\end{equation}
The Lagrange multipliers $\mu$ and $\beta$ are determined by particle and energy conservation,
\begin{equation} \label{eq:consnor}
\int dp\, d\theta f(p,\theta)=1,
\end{equation}
\begin{equation} \label{eq:consene}
\int dp\, d\theta \,f(p,\theta) \,\left[\frac{p^2}{2}-\frac{1}{2}(1-M_x \cos \theta)\right]=u,
\end{equation}
respectively, and the magnetization by the self-consistency requirement, 
\begin{equation} \label{eq:consmag}
\int dp \,d\theta  \cos\theta  f(p,\theta)= M_x.
\end{equation}
Solving these equations numerically along the curve Eq.({\ref{curveLB}}), we see that there is an excellent agreement
between LB theory and the simulations, Fig. 2.  If the macroscopic oscillations are suppressed and 
the parametric resonances are not excited, the system is able to relax to a quasi-ergodic equilibrium permitted by the
Vlasov dynamics.  
\begin{figure} 
\includegraphics[scale=0.8,width=9cm]{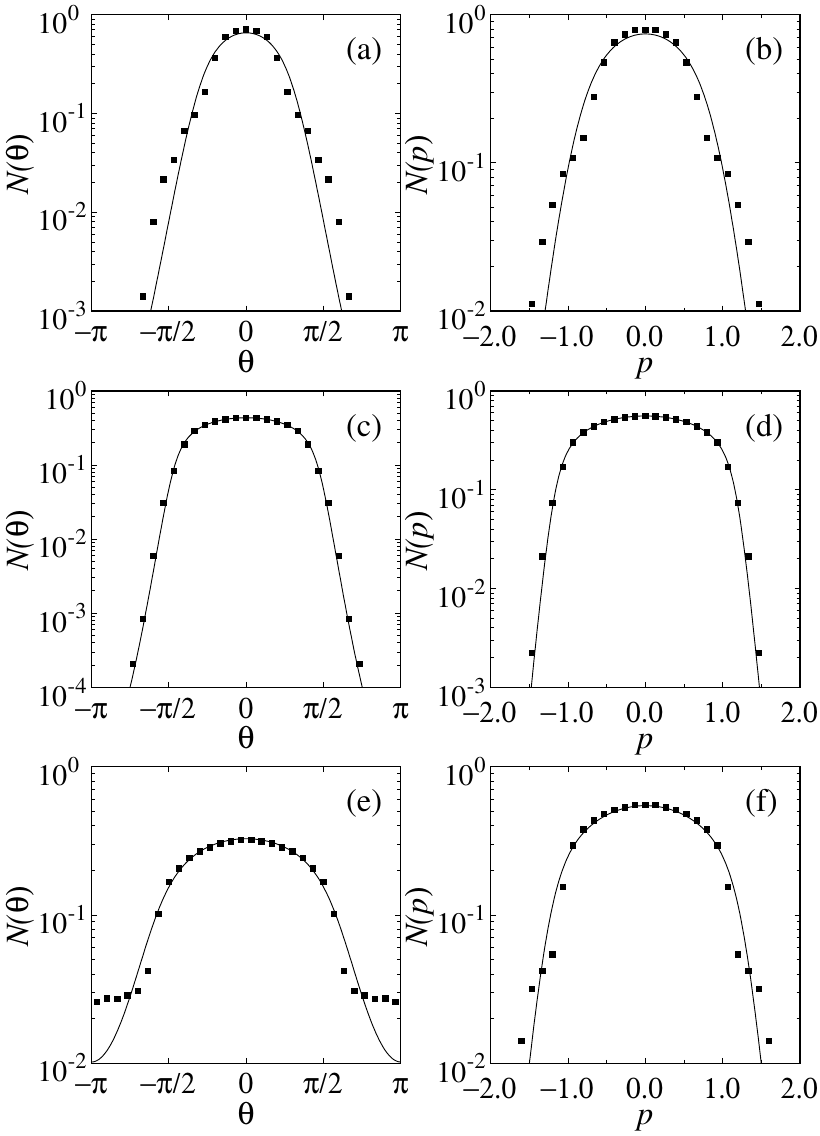}
\caption{The angle  and velocity distribution functions corresponding to the initial conditions described 
by points (A), (B) and (C)  of Fig. 1, respectively.  Symbols are the results of  
molecular dynamics simulations and solid curves are the predictions of  LB theory.  The simulated distribution
functions for the point (B), lying on the generalized virial curve,  are in excellent agreement with the predictions of 
the LB theory [panels (c) and (d)], demonstrating 
that the dynamics along the generalized virial curve is quasi-ergodic. On the other
hand, the distribution functions for points (A) and (C) deviate significantly from the predictions of LB theory  --- showing
that away from the generalized virial curve, ergodicity is broken [panels (a), (b) and (e), (f)].}
\label{fig2}
\end{figure}


To make clear the role
of parametric resonances in ergodicity breaking, in Fig.3a we plot   
the Poincar\'e section of a set of non-interacting test particles, which at $t=0$ are
distributed in accordance with Eq.(\ref{f0}).  The motion of each particle is governed by Eq.(\ref{evol}) 
with $M_x$ determined by Eqs. (\ref{mag}) and (\ref{env}).  The position and momentum  of each particle are 
plotted when magnetization is at its minimum.
We see that if the energy and the initial magnetization lie on the generalized 
virial curve --- point (B) of Fig. 1  ---  particle  trajectories 
are completely regular.  However, when  initial conditions do not coincide with the generalized virial curve --- point (T) of
Fig. 1 --- parametric resonances appear and dynamics becomes chaotic.  Particles enter in resonance with 
the oscillations of the mean-field potential, gaining
sufficient energy to move into statistically improbable regions of the phase space. The  Poincar\'e section 
of test particle dynamics is remarkably similar to the final stationary distribution obtained  using the 
complete N-body molecular dynamics simulation of the HMF, Fig. 3.
Eq. (\ref{env}) can also be used to calculate the period of the first oscillation of $M(t)$. 
For example, for point (T) of the phase diagram Fig. \ref{fig1}, we find the period to be $T=5.0$, while the full molecular dynamics simulations gives $T=5.4$.  For point (C) we find $T=3.85$, while the simulations give $T=3.82$.

\begin{figure} 
\includegraphics[scale=0.8,width=9cm]{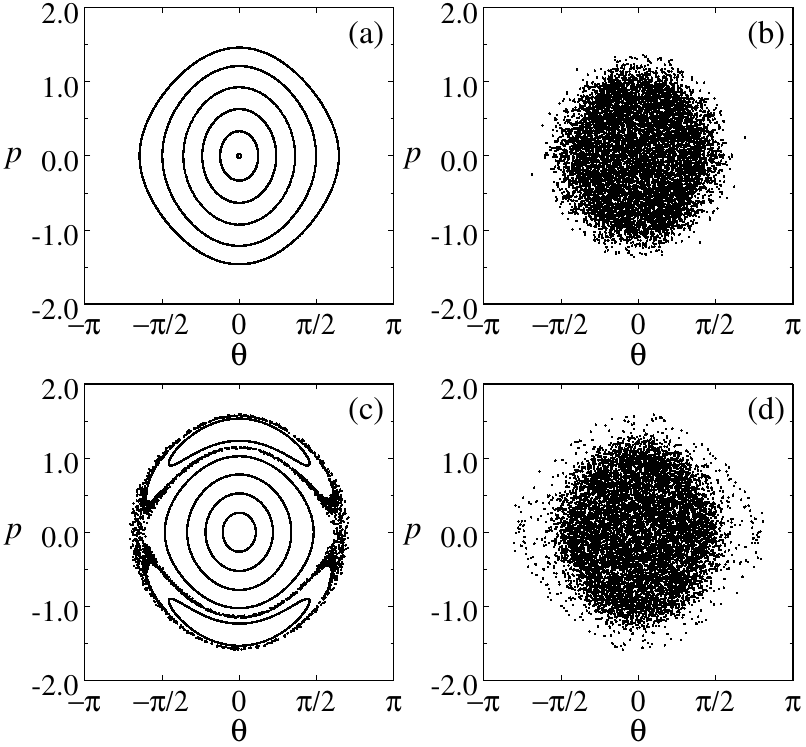}
\caption{Poincar\'e sections of test particles and  snapshots of the phase space obtained using molecular dynamics
simulation once the system has relaxed to qSS. Panels (a) and (b) correspond to the initial condition lying on the 
generalized virial curve, point (B) of Fig. 1.  In this case the test particle dynamics is completely regular,
and the stationary particle distributions are well described by LB theory. 
Panels (c) and (d) correspond to the initial conditions slightly off the virial curve, point (T) of Fig. 1. Even
though we have moved only a little from the virial curve, we see the appearance of resonant islands and
the dynamics of some of the test particles becoming chaotic. Such resonances drive some  particles of the HMF to statistically
improbable --- from the point of view of the Boltzmann-Gibbs and LB statistical mechanics --- 
regions of the phase space. Once the envelope oscillations are damped out, particle dynamics becomes completely integrable,
and there is no mechanism for the resonant particles to equilibrate with the rest of the distribution. Thus, away  
from the generalized
virial curve, ergodicity becomes broken.}
\label{fig3}
\end{figure}

In conclusion, we have studied the mechanism responsible for the 
ergodicity breaking in systems with long-range interactions. Ergodicity breaking and the parametric resonances are
intimately connected.  If the
macroscopic oscillations --- and the resulting resonances are suppressed --- the system is able to relax to a quasi-ergodic
stationary state.  However, when the parametric resonances are excited, some particles are ejected to statistically improbable
regions of the phase space, 
at the same time as the oscillations are damped out.  The process of continuous particle ejection, and the
resulting decrease of  macroscopic oscillations of the envelope, leads to the formation of a static mean-field potential  
and to asymptotically 
integrable dynamics.  
Once the integrability of the equations of motion is achieved, the ergodicity
becomes irreversibly broken.  Unlike for particles with short-range interaction potentials, 
ergodicity is the exception rather than the rule for systems with long-range 
forces --- it can only be observed if the initial distribution function satisfies
the generalized virial condition derived in this Letter. Finally we note, that since the 
 stationary distribution must satisfy the 
virial condition and the energy must be conserved, 
Eq.(\ref{curveLB}) allows us to predict the magnetization to which the system will evolve for 
initial conditions lying inside the
ferromagnetic region.  For example, point (A) of Fig. 1 which has initial magnetization and energy $M_0=0.74$ 
and $u=0.55$, will evolve to a final stationary state with  $M=0.56$; while the point (C) with  $M_0=0.74$ 
and $u=0.25$, will evolve to a final stationary state with  $M=0.86$, which are precisely the values obtained using
the molecular dynamics simulations.

This work was partially supported by the CNPq, FAPERGS, INCT-FCx, and by the US-AFOSR under the grant FA9550-09-1-0283.


\begin{thebibliography}{99}
\bibitem{Campa09}
  A. Campa, T. Dauxois, and S. Ruffo, Phys. Rep. {\bf 480}, 57 (2009).
\bibitem{Gibbs} 
  J. W. Gibbs, {\it Collected Works}, Longmans, Green and Co., NY (1928).
\bibitem{Levinprl}
  Y. Levin, R. Pakter, and T. N. Teles, Phys. Rev. Lett. {\bf 100}, 040604 (2008).
\bibitem{TaLe10} T.N. Teles, Y.Levin, R. Pakter, and   F.B. Rizzato, J. Stat. Mech. P05007 (2010).
\bibitem{JoWo} M. Joyce and T. Worrakitpoonpon, Phys. Rev. E 84, 011139 (2011).
\bibitem{Tel2011} Teles, T. N. and Levin, Y. and Pakter, R., Mon. Not. R. Astron. Soc. {\bf 417}, 1, (2011).
\bibitem{Barre2001} J. Barr{\'e}, D. Mukamel, and S. Ruffo, Phys. Rev. Lett. {\bf 87}, 030601 (2001).
\bibitem{Kav2007} K. Jain et al., J. Stat. Mech. (2007) P11008.
\bibitem{Ly67} D. Lynden-Bell, Mon. Not. R. Astron. Soc. {\bf 136}, 101 (1967).
\bibitem{Br77} W. Braun and K. Hepp, Comm. Math. Phys.  {\bf 56}, 101 (1977).
\bibitem{AnCa07} A. Antoniazzi, F. Califano, D. Fanelli, and S. Ruffo, Phys. Rev. Lett. {\bf 98}. 150602 (2007);
 J. Barre, F. Bouchet, T. Dauxois, S. Ruffo, Phys. Rev. Lett. {\bf 89}, 110601 (2002); T. M. Rocha Filho, A.  Figueiredo,  M. A.  Amato, Phys. Rev. Lett. {\bf 95}. 190601 (2005).
\bibitem{AnFa07} A. Antoniazzi, D. Fanelli, and S. Ruffo, Y.Y. Yamaguchi Phys. Rev. Lett. {\bf 99}. 040601 (2007).
\bibitem{AnFa07b} A. Antoniazzi, D. Fanelli, J. Barr\'e.,  P.H. Chavanis, T. Dauxois
and S. Ruffo, Phys. Rev. E {\bf 75}, 011112  (2007);
\bibitem{Muka2005}
  D. Mukamel,   and S. Ruffo,   and N. Schreiber, Phys. Rev. Lett. {\bf 95}, 240604 (2005).
\bibitem{La46} L. Landau, J. Phys. USSR {\bf 10}, 25  (1946).
\bibitem{BaCh08} R. Bachelard, C. Chandre,D. Fanelli, X. Leoncini, and S. 
Ruffo,  Phys. Rev. Lett. {\bf 101}. 260603 (2008).
\bibitem{PaLe11} R. Pakter and Y. Levin, Phys. Rev. Lett.  {\bf 106}, 200603 (2011). 
\bibitem{StCh09} F. Staniscia, P.H. Chavanis, G. De Ninno, D. Fanelli,
Phys. Rev. E  {\bf 80}, 021138  (2009).




\end{thebibliography}
\end{document}